\documentstyle[NATO,epsf]{CRCKAPB}

\begin{opening}
\title{Zeeman tomography of magnetic white dwarfs:\protect\\
General method and application to EF\,Eridani} 

\author{F.~Euchner} 
\author{K.~Beuermann} 
\author{K.~Reinsch} 
\author{S.~Jordan}
\author{F.V.~Hessman}
\institute{Universit\"ats-Sternwarte G\"ottingen, Geismarlandstr.~11,\\ 
D-37083 G\"ottingen, Germany\footnote{SJ: now at Institut f\"ur Astronomie und
Astrophysik, Universit\"at T\"ubingen, Germany}} 
\author{B.T.~G\"ansicke} \institute{Dept. of
Physics \& Astronomy, University of Southampton, Southampton SO17 1BJ,
UK}
\end{opening}

\begin{document}

\begin{abstract} 
We have developed a new method to derive the magnetic field
distribution on the surfaces of rotating magnetic white dwarfs from
phase-resolved flux and circular polarization spectra. An optimization
code based on an evolutionary strategy is used to fit synthetic Zeeman
spectra for a variety of model geometries described in the framework
of a truncated multipole expansion.  We demonstrate that the code
allows the reconstruction of relatively complex fields using
noise-added synthetic input spectra.  As a first application, we
analyze flux and circular polarization spectra of the polar EF\,Eri in
a low state of accretion taken with FORS1 at the ESO VLT.
\end{abstract}

\section{Reconstruction of the field geometry from synthetic spectra}
We use a pre-computed database of 46\,800 synthetic flux and
polarization spectra for field strengths of 1--400\,MG and a wide
range of effective temperatures to generate synthetic Zeeman spectra,
which we subject to an automatic optimization scheme with the aim to
recover the parameters describing the original magnetic field
configuration.  We adopt input field geometries which involve the sum
of non-aligned dipole and quadrupole components. We also provide for
off-centre shifts of the configuration.  By adding Gaussian noise we
create input spectra with \mbox{$S/N$ = 100} and 20, respectively.

\begin{figure}[t]
\centering
\epsfxsize=0.80\textwidth \epsfbox{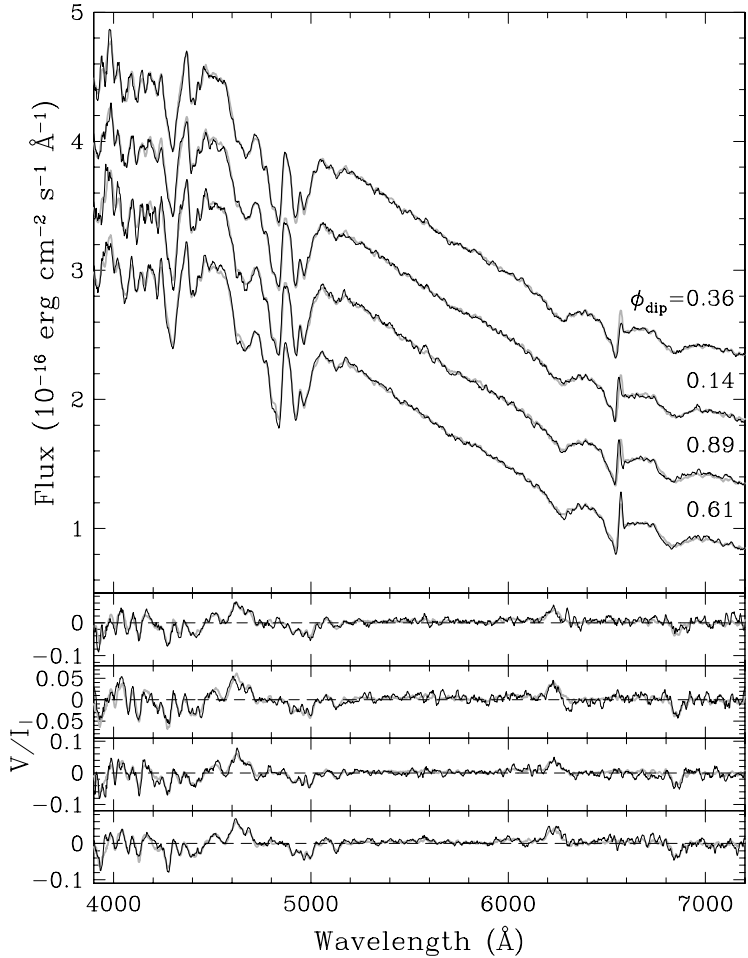}
\caption{Flux {\it (top)} and circular polarization {\it (bottom)}
spectra of EF\,Eri, taken on 22 Nov 2000 with FORS1 at the ESO VLT
({\it black:} spectrum at phase $\phi$, {\it grey:} orbital mean
spectrum). The upper three flux spectra have been shifted vertically
to avoid overlap.}
\end{figure}

Off-centred dipole-quadrupole combinations are reliably recovered even
for \mbox{$S/N = 20$} by our code, which can handle up to 12 free
parameters but becomes inefficient if higher multipoles are
included. The poor convergence is caused by the complexity of the
$\chi^2$-landscape, which develops an increasing number of secondary
minima for more complex field geometries (for a detailed
investigation, see Euchner et al., 2002). We find that a given set of
phase-resolved Zeeman spectra can be reproduced within the noise by
quite different formal representations of the field, which, however,
all seem to describe rather similar actual field geometries. We are
confident, therefore, that observed phase-resolved Zeeman spectra can
provide quite definite information on the field structure.

\section{The magnetic field of EF\,Eridani}

\begin{figure}[t]
\centering
\epsfxsize=0.85\textwidth \epsfbox{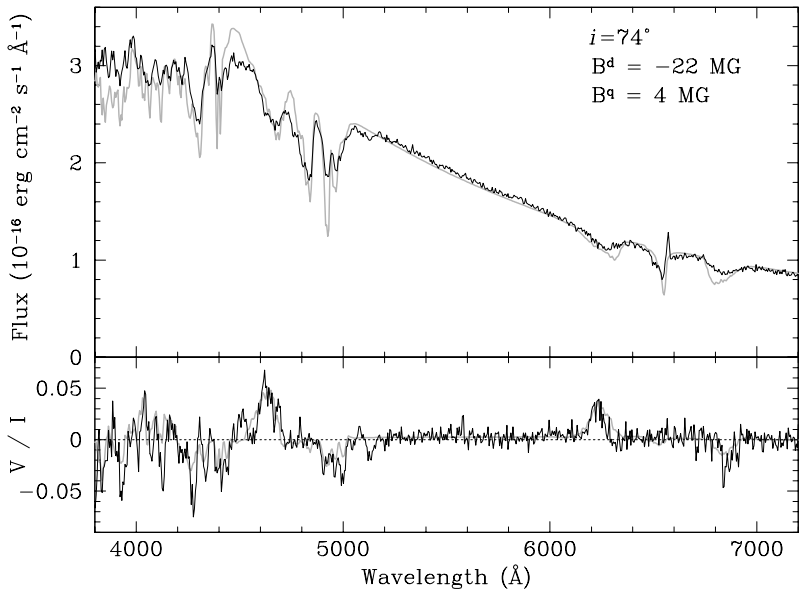}
\caption{Dipole+quadrupole fit to the orbital mean flux and circular
polarization spectra of EF\,Eri ({\it black:} VLT observation from
Nov 2000, {\it grey:} model).}
\end{figure}

The magnetic field structure over the surface of accreting white
dwarfs in cataclysmic binaries becomes accessible to observation only
in states of low or switched-off accretion. Following the first
detection of Zeeman lines in the polar EF\,Eridani ($P_{\rm orb}$ = 81
min) by Wheatley \& Ramsay (1998), we obtained phase-resolved flux and
circular polarization spectra of this object with FORS1 at the ESO VLT
on 22 Nov 2000. Fig.\,1 shows the spectra collected in four almost
equal phase bins after removal of a slight rotational temperature
variation. There is no obvious variation of the Zeeman structures with
phase. Fits of inclined multipole field models with the angle $\beta$
between magnetic and rotational axis left free yielded \mbox{$\beta
\simeq$~0}. The lack of phase dependence of the Zeeman features
further suggests that the tesseral/sectoral multipole components are
weak. We fitted the mean flux and polarization spectra, therefore,
with an expansion including only the zonal \mbox{($m = 0$)} components
up to a maximum degree \mbox{$l = l_{\rm max}$}.

Fig.\,2 shows the dipole+quadrupole fit \mbox{($l_{\rm max} =2$)},
which is substantially better than the pure dipole (not
shown). Fig.\,3 depicts the result for the expansion up to
\mbox{$l_{\rm max} = 5$}. The dipole is replaced by large coefficients
for the higher-order multipole components which substantially improve
the fit. While the higher-order multipole fit accounts for many of the
observed details -- all the wiggles in the blue are genuine Zeeman
features -- there are still noticeable discrepancies: (i) the observed
lack of circular polarization around 5700\,\AA\ requires that a larger
fraction of the \mbox{$B \sim$~40\,MG} field occurs at negative $\cos
\psi$ (angle between line of sight and field direction), and (ii) the
observed narrow spectral feature at 5120\,\AA\ can be fitted by
additional field contributions with $B$ between 60 and 100\,MG
pointing away from the observer, possibly generated by a steeper rise
of $B$ towards the lower (unseen) pole than provided by the model.

We conclude that the magnetic field of EF\,Eri deviates strongly from
a centred dipole, but do not pretend to have arrived at a definitive
solution for its field structure.

\begin{figure}[t]
\centering
\epsfxsize=0.82\textwidth \epsfbox{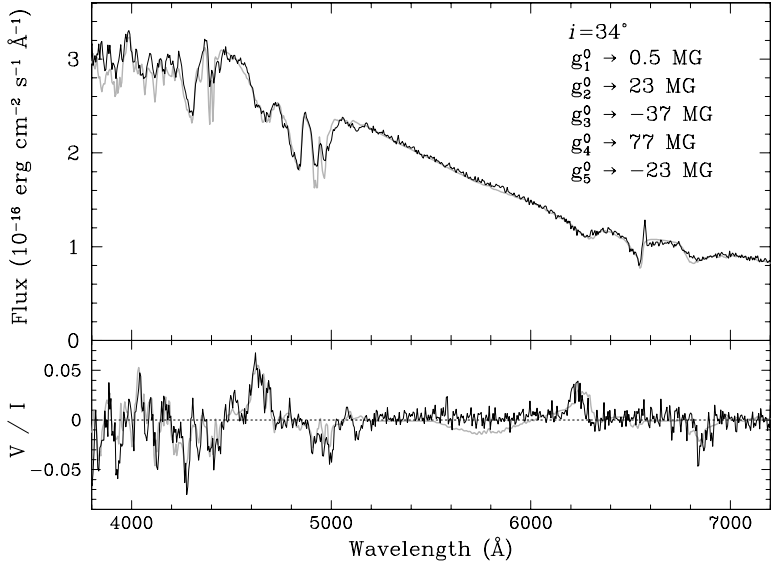}
\caption{Higher-order multipole fit to the orbital mean flux and
circular polarization spectra of EF\,Eri ({\it black:} VLT observation
from Nov 2000, {\it grey:} model).}
\end{figure}

\end{document}